\documentclass{article}  
\usepackage{breckenr2005}
\usepackage{graphicx}
\frompage{000} \topage{000}                                              %|
\def\rightmark{Isospin Asymmetry}
\def\leftmark{A.~W. Steiner, et. al.}

\title{Isospin Asymmetry in Nuclei, Neutron Stars, and Heavy-Ion 
Collisions}
\authors{
{A.W. Steiner$^1$, M. Prakash$^2$, J.M. Lattimer$^2$, and P.J. Ellis$^{3}$ %
}\\[2.812mm]
{\normalsize
\hspace*{-8pt}$^1$ Theoretical Division, Los Alamos National
Laboratory,\\ Los Alamos, NM 87545, USA \\
\hspace*{-8pt}$^2$ Department of Physics and Astronomy, SUNY at Stony
Brook,\\ Stony Brook, New York 11794-3800, USA \\
\hspace*{-8pt}$^3$ School of Physics and Astronomy, University of
Minnesota,\\ Minneapolis, MN 55455-0112, USA \\
}}
 
\abstract{The roles of isospin asymmetry in nuclei and neutron stars
are investigated using a range of potential and field-theoretical
models of nucleonic matter. The parameters of these models are fixed
by fitting the properties of homogeneous bulk matter and closed-shell
nuclei. We discuss and unravel the causes of correlations among the
neutron skin thickness in heavy nuclei, the pressure of
beta-equilibrated matter at a density of 0.1 fm$^{-3}$, and the radii
of moderate mass neutron stars.  The influence of symmetry energy on
observables in heavy-ion collisions is summarized. }

 
\begin{document}
 
\maketitle
\setcounter{page}{1}

\section{The Symmetry Energy}\label{s:defn}

The nuclear symmetry energy is the energy required to create an
asymmetry between neutrons and protons (i.e., an isospin
asymmetry). The preference for equal numbers of neutrons and protons
is manifest in the difference between the deuteron, which is bound,
and the dineutron, which is not. In nuclei, the preferred energy state
would have equal numbers of neutrons and protons if the Coulomb
interaction was absent.  Expressed in terms of the energy density of
homogeneous matter, the symmetry energy is given by
\begin{equation}
E_{sym}(n) = \left. \left( \frac{1}{2} 
\frac{d^2 (\varepsilon/n)}{ d \delta^2}\right) \right|_{n,\delta=0} \, ,
\end{equation}
where $n$ is the baryon density, $\delta=(n_n-n_p)/n$ is the neutron-to-proton
asymmetry, $n_n$ and $n_p$ are the neutron and proton
densities, and $\varepsilon$ is the energy density.

\begin{figure}[htb]
\begin{center}
\includegraphics[scale=0.63,angle=0]{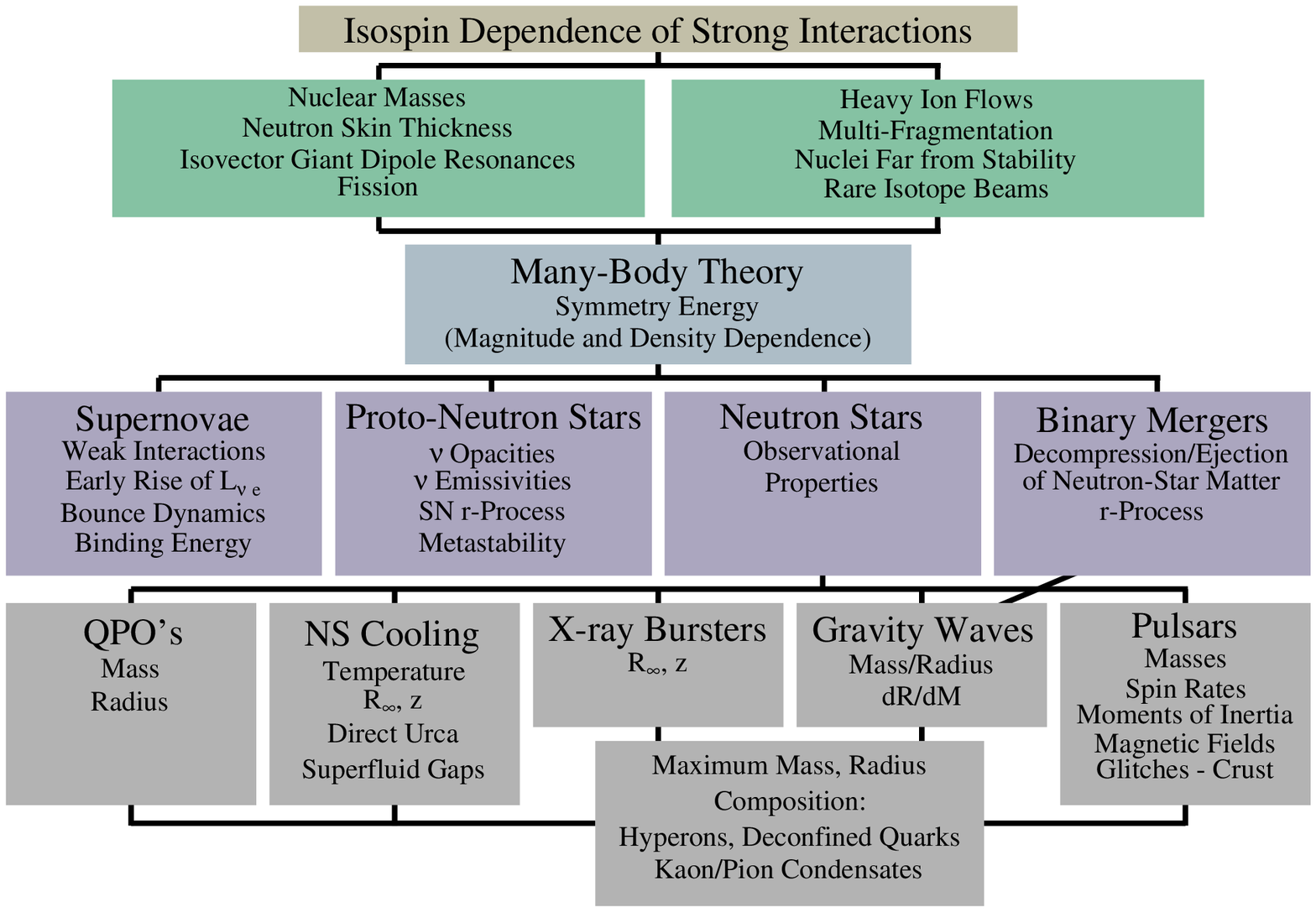}
\end{center}
\vspace*{-0.8cm}
\caption{The multifaceted influence of the nuclear 
symmetry energy.}
\label{f:corpdiag}
\vspace*{-0.6cm}
\end{figure}

The symmetry energy influences several aspects of nuclear physics, from
giant dipole resonances to heavy-ion collisions, and also several
astrophysical processes, from supernovae to neutron stars. The broad
influence of the symmetry energy is illustrated in
Fig.~\ref{f:corpdiag}~\cite{Steiner05}.
  
\begin{figure}[htb]
\vspace*{-0.3cm}
\begin{center}
\includegraphics[scale=0.45,angle=0]{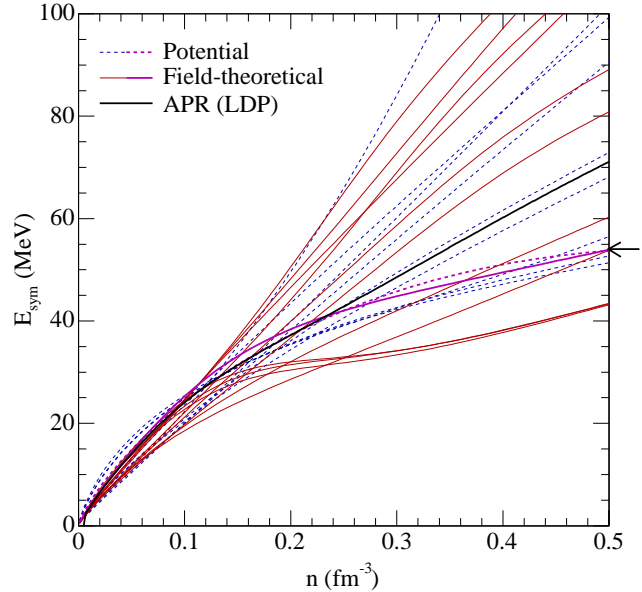}
\end{center}
\vspace*{-0.9cm}
\caption{The symmetry energy versus density for
various equations of state.  Solid (dashed) lines are for
field-theoretical (potential) models. The thick solid line shows
the APR symmetry energy for the low density phase.
The arrow identifies the NRAPR (dotted line) and
RAPR (solid line) models which have nearly identical
symmetry energies at $n=0.5$ fm$^{-3}$. }
\vspace*{-0.8cm}
\label{f:syme}
\end{figure}

In spite of the diverse influences of the symmetry energy, its
magnitude and density dependence are not well
understood. Figure~\ref{f:syme} shows the symmetry energy as a
function of density for the models considered in this work (details
are discussed in Section~\ref{s:eos}).  The importance of the symmetry
energy and our relative ignorance of it motivates studies of nuclear
and astrophysical observables that relate to the symmetry energy, and
of how knowledge about the symmetry energy can help predict the outcome
of experiments and astronomical observations.

\section{The Equation of State}
\label{s:eos}  

We will consider field-theoretical models in which a Walecka-type
Lagrangian is utilized in the mean field approximation, and
potential-models in which effective zero-range forces are used to
construct an effective Hamiltonian density. In addition to these two
general classes, we also employ the equation of state of Akmal,
Pandharipande, and Ravenhall~\cite{Akmal98} (APR). Because first
principle Monte Carlo calculations of the structure of heavy nuclei
are not yet available, we also perform field-theoretical (RAPR) and
potential-model (NRAPR) fits to the APR EOS. Details of these fits are
given in Ref.~\cite{Steiner05}.

We require that all models reproduce 
the properties of nuclear matter at saturation density: 
\begin{eqnarray}
{\rm equilibrium~binding~energy}~: B &=& -16\pm 1~{\rm MeV}\,,
\nonumber \\
{\rm equilibrium~density}~: n_0 &=&
0.16\pm 0.01~{\rm fm}^{-3}\,,
\nonumber \\
{\rm incompressibility}~: K&=&(200-300)~{\rm  MeV}\,,
\nonumber \\
{\rm Landau~effective~mass}~: m_L^* &=& (0.6-1.0)~M \,, \quad {\rm and}
\nonumber \\
{\rm symmetry~energy}~: S_v&=& (25-35)~{\rm MeV}\,.
\end{eqnarray}

In the case of potential models, calculations of nuclei are performed
using the Hartree-Fock-Bogoliubov approach \cite{Reinhard99} that
includes pairing interactions. Hartree calculations \cite{Horowitz81}
are employed in the field-theoretical approach, as a treatment of
the exchange (Fock) terms is considerably more complicated than in the
potential model approach.  In both approaches, we require that the
binding energy and the charge radii of closed-shell nuclei are
reproduced to within $2\%$ of the measured values.  The scalar meson
mass in the field theoretical approach was restricted to lie between
450 and 550 MeV. 

In all cases considered the supranuclear EOS was constrained to
yield a maximum neutron star mass of at least $1.44{\rm M}_\odot$, the
larger of the accurately measured neutron star masses in the double
neutron star binary PSR1913+16 (see Ref.~\cite{Lattimer04} for a
compilation of known masses). 

\begin{figure}[htb]
\begin{center}
\includegraphics[scale=0.40,angle=0]{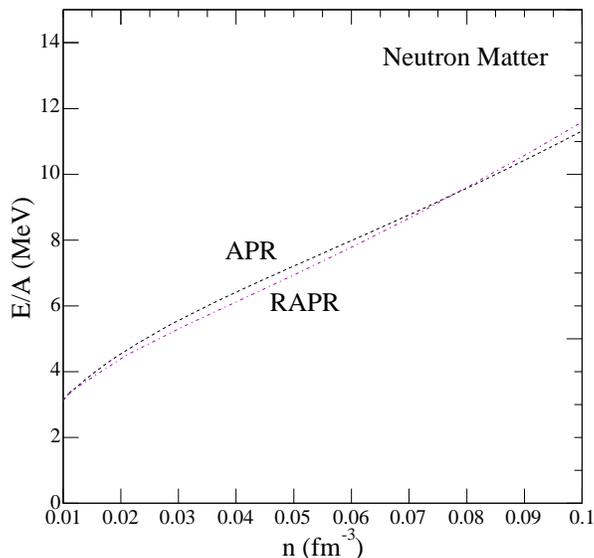}
\end{center}
\vspace*{-0.9cm}
\caption{The energy per baryon of neutron matter in the 
APR and RAPR models. The energies in both models are approximately 
equal to half of the Fermi gas energies.}
\label{f:eneut}
\vspace*{-0.8cm}
\end{figure}

Monte Carlo calculations of low-density neutron matter have shown that
the energy per baryon should be approximately half the Fermi
energy~\cite{Carlson03}. The APR equation of state, 
also based on Monte Carlo calculations, exhibits
this behavior. By fitting the parameters of our field-theoretical
model to APR, we can obtain this kind of behavior in RAPR as
demonstrated in Fig.~\ref{f:eneut} where the energies per baryon of APR
and RAPR are compared.

\section{Neutron Stars and Nuclei}
\label{s:correl}

Recently, several empirical relationships have been discovered that
underscore the role of isospin interactions in nuclei and neutron
stars. 

\begin{figure}[htb]
\vspace*{-0.2cm}
\begin{center}
\includegraphics[scale=0.40,angle=0]{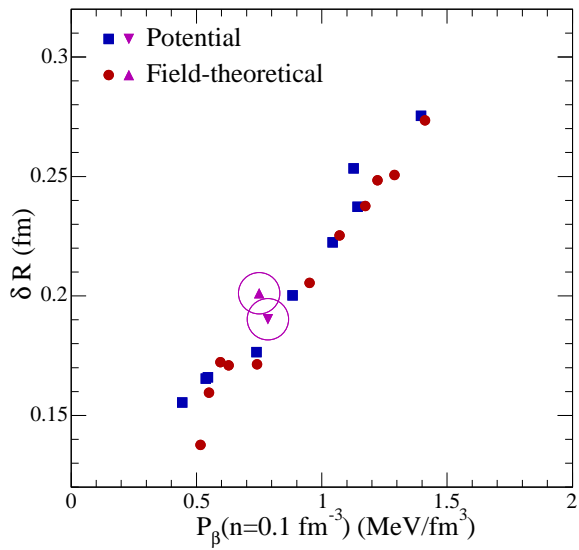}
\end{center}
\vspace*{-0.9cm}
\caption{The neutron skin thickness $\delta R$ of finite nuclei versus
the pressure of $\beta$-equilibrated matter at a density of 0.1
fm$^{-3}$. The blue squares display the results for potential models
and the red circles show the results for field-theoretical modes.
The circled triangles represent the potential (NRAPR) and
field-theoretical (RAPR) model fits to APR. }
\label{f:typel}
\vspace*{-0.4cm}
\end{figure}

{\em The neutron skin thickness in nuclei is correlated to the pressure of
pure neutron matter at sub-nuclear density}~: Typel and Brown
\cite{Brown00,Typel01} have noted that model calculations of the
difference between neutron and proton rms radii $\delta R ={\langle
r_n^2\rangle}^{1/2} - {\langle r_p^2\rangle}^{1/2}$ are linearly
correlated with the pressure of pure neutron matter at a density 
characteristic of the mean density in the nuclear surface (e.g.,
0.1 fm$^{-3})$. We show this correlation in Fig.~\ref{f:typel} for the
models considered in this work.  The density dependence of the
symmetry energy controls $\delta R$ (termed the neutron
skin thickness) in a heavy nucleus. $\delta
R$ is proportional to a specific average of
$[1-E_{sym}(n_0)/E_{sym}(n)]$ in the nuclear surface
~\cite{Krivine84,Lattimer96,Steiner05}.  To the extent that this
correlation is valid, a measurement of $\delta R$ will help 
establish an empirical calibration point for the pressure of neutron
star matter at subnuclear densities.

\begin{figure}[htb]
\vspace*{-0.3cm}
\begin{center}
\includegraphics[scale=0.50,angle=0]{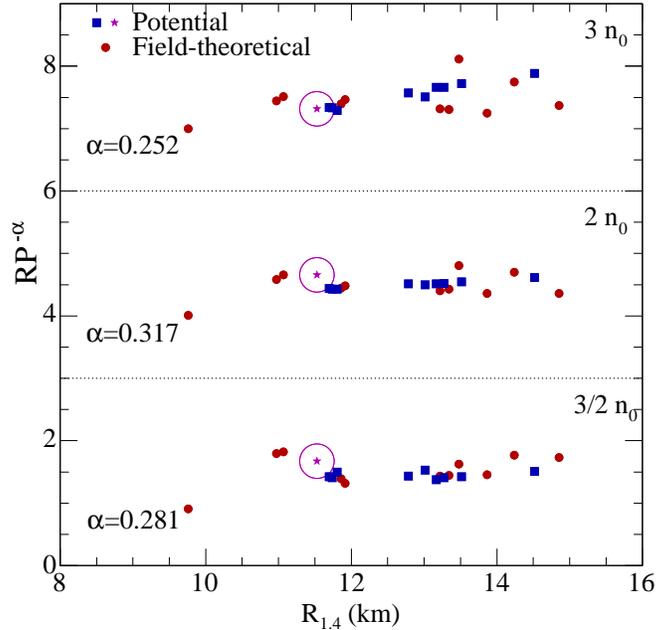}
\end{center}
\vspace*{-0.8cm}
\caption{The quantity $RP^{-\alpha}$ as a function of the radius $R$
of a $1.4M_{\odot}$ star for pressures $P$ determined at 3/2, 2 and 3
times equilibrium nuclear matter density.  For each density, the
best-fit value for the exponent $\alpha$ is as indicated.  Circled
stars indicate the results obtained with the APR equation of state.}
\label{f:latpi}
\vspace*{-0.9cm}
\end{figure}

{\em The neutron star radius $R$ and the pressure $P$ of neutron-star
matter}~: Lattimer and Prakash \cite{Lattimer00,Lattimer01} found that
the quantity $RP^{-1/4}$ is approximately constant, for a given
neutron star mass, for a wide variety of equations of state when the
pressure $P$ of beta-equilibrated neutron-star matter is evaluated at
a density in the range $n_0$ to $2n_0$, where $n_0$ denotes
equilibrium nuclear matter density. Since the pressure of nearly pure
neutron matter (a good approximation to neutron star matter) near
$n_0$ is approximately given by $n^2\partial E_{sym}/\partial n$, the
density dependence of the symmetry energy just above $n_0$ determines
the neutron star radius. Figure~\ref{f:latpi} shows this correlation
as $RP^{-\alpha}$ versus $R$ for stars of mass $1.4{\rm M}_\odot$ for
the EOS's considered here and densities $n=1.5-3 n_0$.  In each case,
the exponent $\alpha$ was determined by a least-squares analysis.

\section{Heavy-Ion Collisions}

\subsection{Multi-fragmentation}

The breakup of excited nuclei into several smaller fragments during an
in\-ter\-med\-i\-ate-energy 
heavy-ion collision probes the phase diagram of
nucleonic matter at sub-saturation density and moderate ($\sim 10-20$
MeV) temperatures. In this region of the phase diagram the system is
mechanically unstable if $(dP/dn)_{T,x}<0$, and/or chemically
unstable if $(d\mu_p/d x)_{T,P}<0$. (A pedagogical account of 
such instabilities can be found in
Ref.~\cite{Chomaz04}). These instabilities, which are directly related
to the symmetry energy at sub-saturation densities~\cite{Li97}, are
believed to trigger the onset  of multifragmentation. Because of the
instabilities, matter separates into coexisting liquid and gas phases,
which each have different proton fractions, {\it i.e.} ``isospin
fractionation''~\cite{Xu00}.  This fractionation is observed in the
isotopic yields which can potentially reveal information about the symmetry
energy.  Also, the scaling behavior of ratios of isotope yields  
measured in separate nuclear reactions, ``isoscaling'', is sensitive
to the symmetry energy~\cite{Tsang01,Ono03}. 
To date, there are many suggestions of
how the symmetry energy may affect multifragmentation~\cite{Das04}. Ongoing
research is concerned with an extraction of reliable constraints on
the symmetry energy from the presently available experimental
information. See also the work by Bao-An Li in this volume.

\subsection{Collective Flow}

Nuclear collisions in the range $E_{lab}/A = 0.5-2$ GeV offer the
possibility of pinning down the equation of state of matter above
normal nuclear density (up to $\sim 2 ~{\rm to}~ 3n_0$) from a study
of matter, momentum, and energy flow of nucleons \cite{Gutbrod89}.
The observables confronted with theoretical analyses include (i) the
mean transverse momentum per nucleon $\langle p_x \rangle /A$ versus
rapidity $y/y_{proj}$ \cite{Danielewicz85}, (ii) flow angle from a
sphericity analysis \cite{Gustafsson84}, (iii) azimuthal distributions
\cite{Welke88}, and (iv) radial flow \cite{Siemens79}. Flow data
to date are largely for protons (as detection of neutrons is
more difficult) and for collisions of laboratory nuclei in which the
isospin asymmetry is not large. 

The prospects of rare iso\-tope accelerators (RIA's) that can collide
high\-ly neu\-tron-rich nuclei has spurred further work to study a system
of neutrons and protons at high neutron
excess~\cite{Das03,Li04,Li04b}.  Observables that are expected to shed
light on the influence of isospin asymmetry include neutron-proton
differential flow and the ratio of free neutron to proton multiplicity
as a function of transverse momentum at midrapidity.  Experimental
investigations of these signatures await the development of RIA's at
GeV energies.  In this connection, it will be important to detect
neutrons in addition to protons.

\section{Outlook}
\label{s:concl}

The correlations presented here will (with more progress in
theory and experiement) help to determine the symmetry energy and
its density dependence. It will be interesting to explore further
connections involving isospin asymmetry in heavy-ion observables and
observables in nuclear structure and astrophysics. Work on this front
is in progress~\cite{Steiner05b}.
 
\section*{Acknowledgment(s)}
Research support of the U.S. Department of Energy under grant numbers
DOE/W-7405-ENG-36 (for AWS), DOE/DE-FG02-87ER-40317 (for MP and JML),
and DOE/DE-FG02-87ER-40328 (for PJE and AWS) is acknowledged. We
thank Chuck Horowitz and David Dean for providing us with computer
codes for the calculation of the properties of finite nuclei.

\bibliography{proc3}

\begin{thebibliography}{10}

\bibitem{Steiner05}
A.~W. Steiner, M.~Prakash, J.~M. Lattimer, and P.~J. Ellis,
\newblock {\em Phys. Rep.} {\bf 411} (2005) 325.

\bibitem{Akmal98}
A.~Akmal, V.~R. Pandharipande, and D.~G. Ravenhall,
\newblock {\em Phys. Rev.} {\bf C58} (1998) 1804.

\bibitem{Reinhard99}
P.-G. Reinhard, D.~J. Dean, W.~Nazarewicz, J.~Dobaczewski, J.~A. Maruhn, and
  M.~R. Strayer,
\newblock {\em Phys. Rev.} {\bf C60} (1999) 014316.

\bibitem{Horowitz81}
C.~J. Horowitz and B.~D. Serot,
\newblock {\em Nucl. Phys.} {\bf A368} (1981) 503.

\bibitem{Lattimer04}
J.~M. Lattimer and M.~Prakash,
\newblock {\em Science} {\bf 304} (2004) 536.

\bibitem{Carlson03}
J.~Carlson, J.~Morales~Jr., V.~R. Pandharipande, and D.~G. Ravenhall,
\newblock {\em Phys. Rev.} {\bf C68} (2003) 025802.

\bibitem{Brown00}
B.~A. Brown,
\newblock {\em Phys. Rev. Lett.} {\bf 85} (2000) 5296.

\bibitem{Typel01}
S.~Typel and B.~A. Brown,
\newblock {\em Phys. Rev.} {\bf C64} (2001) 027302.

\bibitem{Krivine84}
H.~Krivine,
\newblock {\em Jour. de Phys. Supp.} {\bf C6} (1984) 153.

\bibitem{Lattimer96}
J.~M. Lattimer,
\newblock {\em Nuclear Equation of State,} eds A. Ansari and
L. Satpathy, World Scientific, Singapore, 1996, p. 83.

\bibitem{Lattimer00}
J.~M. Lattimer and M.~Prakash,
\newblock {\em Phys. Rep.} {\bf 333} (2000) 121.

\bibitem{Lattimer01}
J.~M. Lattimer and M.~Prakash,
\newblock {\em Astrophys. J.} {\bf 550} (2001) 426.

\bibitem{Chomaz04}
P.~Chomaz, M.~Colonna, and J.~Randrup,
\newblock {\em Phys. Rep.} {\bf 389} (2004) 263.

\bibitem{Li97}
B.~A. Li and C.~M. Ko,
\newblock {\em Nucl. Phys.} {\bf A618} (1997) 498.

\bibitem{Xu00}
H.~S. Xu, M.~B. Tsang, T.~X. Liu, X.~D. Liu, W.~G. Lynch, et. al.,
\newblock {\em Phys. Rev. Lett.} {\bf 85} (2000) 716.

\bibitem{Tsang01}
M.~B. Tsang, W.~A. Friedman, C.~K. Gelbke, W.~G. Lynch, G.~Verde, et. al.,
\newblock {\em Phys. Rev. Lett.} {\bf 86} (2001) 5023.

\bibitem{Ono03}
A.~Ono, P.~Danielewicz, W.~A. Friedman, W.~G. Lynch, and M.~B. Tsang,
\newblock {\em Phys. Rev.} {\bf C68} (2003) 051601(R).

\bibitem{Das04}
C.~B. Das, S.~Das~Gupta, W.~G. Lynch, A.~Z. Mekjian, and M.~B. Tsang,
\newblock {\em Phys. Rep.} {\bf 406} (2005) 1.

\bibitem{Gutbrod89}
H.~H. Gutbrod, A.~M. Poskanzer, and H.~G. Ritter,
\newblock {\em Rep. Prog. Phys.} {\bf 52} (1989) 267.

\bibitem{Danielewicz85}
P.~Danielewicz and G.~Odyniec,
\newblock {\em Phys. Lett.} {\bf B157} (1985) 146.

\bibitem{Gustafsson84}
H.~A. Gustafsson, H.~H. Gutbrod, B.~Kolb, H.~L\"{o}hner, B.~Ludewigt, et. al.,
\newblock {\em Phys. Rev. Lett.} {\bf 52} (1984) 1590.

\bibitem{Welke88}
G.~M. Welke, M.~Prakash, T.~T.~S. Kuo, S.~Das~Gupta, and C.~Gale,
\newblock {\em Phys. Rev.} {\bf C38} (1988) 2101.

\bibitem{Siemens79}
P.~Siemens and J.~O. Rasmussen,
\newblock {\em Phys. Rev. Lett.} {\bf 42} (1979) 880.

\bibitem{Das03}
C.~B. Das, S.~Das~Gupta, C.~Gale, and B.-A. Li,
\newblock {\em Phys. Rev.} {\bf C67} (2003) 034611.

\bibitem{Li04}
B.-A. Li, C.~B. Das, S.~Das~Gupta, and C.~Gale,
\newblock {\em Phys. Rev.} {\bf C88} (2004) 192701.

\bibitem{Li04b}
B.-A. Li, C.~B. Das, S.~Das~Gupta, and C.~Gale,
\newblock {\em Nucl. Phys.} {\bf A735} (2004) 563.

\bibitem{Steiner05b}
A.~W. Steiner, B.-A. Li,
\newblock nucl--th/0505051, 2005.

\end{thebibliography}
\bibliographystyle{unsrt}

\vfill\eject
\end{document}